# Paired spiking robustly shapes spontaneous activity in neural networks *in vitro*

Aurel Vasile Martiniuc[1]*, Victor Bocoş-Binţinţan[2], Rouhollah Habibey[3], Asiyeh Golabchi[3], Alois Knoll[1], Axel Blau[3]

[1] Computer Science Department VI, Technical University Munich, Boltzmannstraße 3, 85748 Garching, Germany, http://www.in.tum.de/
[2] Dept. of Environmental Science & Engineering, Babeş-Bolyai University, 30 Fantanele Street, 400294 Cluj-Napoca, Romania, http://enviro.ubbcluj.ro/
[3] Dept. of Neuroscience and Brain Technologies, Italian Institute of Technology, via Morego 30, 16163 Genoa, Italy, http://www.iit.it

*Correspondence: martiniv@in.tum.de

**Abstract**

*In vivo*, neurons establish functional connections and preserve information along their synaptic pathways from one information processing stage to the next in a very efficient manner. Paired spiking (PS) enhancement plays a key role by acting as a temporal filter that deletes less informative spikes. We analyzed the spontaneous neural activity evolution in a hippocampal and a cortical network over several weeks exploring whether the same PS coding mechanism appears in neuronal cultures as well. We show that self-organized neural *in vitro* networks not only develop characteristic bursting activity, but feature robust *in vivo*-like PS activity. PS activity formed spatiotemporal patterns that started at early days *in vitro* (DIVs) and lasted until the end of the recording sessions. Initially random-like and sparse PS patterns became robust after three weeks *in vitro* (WIVs). They were characterized by a high number of occurrences and short inter-paired spike intervals (IPSIs). Spatially, the degree of complexity increased by recruiting new neighboring sites in PS as a culture matured. Moreover, PS activity participated in establishing functional connectivity between different sites within the developing network. Employing transfer entropy (TE) as an information transfer measure, we show that PS activity is robustly involved in establishing effective connectivities. Spiking activity at both individual sites and network level robustly followed each PS within a short time interval. PS may thus be considered a spiking predictor. These findings suggest that PS activity is preserved in spontaneously active *in vitro* networks as part of a robust coding mechanism as encountered *in vivo*. We suggest that, presumably in lack of any external sensory stimuli, PS may act as an internal surrogate stimulus to drive neural activity at different developmental stages.

Martiniuc *et al.*                    Paired spiking robustly shapes spontaneous activity in neural networks *in vitro*

## 1. Introduction

Both *in vivo* and *in vitro,* synchronous correlated activity known as bursting is one of the information processing mechanisms that shape network interconnectivity, both at single cell and network level (van Pelt *et al.,* 2004; Wagenaar *et al.,* 2005; McCabe *et al.,* 2006; Wagenaar *et al.,* 2006; Rolston *et al.,* 2007; Mazzoni *et al.,* 2007; Sun *et al.,* 2010).

Bursting not only occurs in brain slices with partially intact interconnectivity (Blankenship and Feller 2010; Rolston *et al.,* 2007), but is also found in neural cultures derived from dissociated brain tissue where it becomes predominant as cultures mature (Wagenaar *et al.,* 2005; Wagenaar *et al.,* 2006). Bursting activity varies with culture age (Nadasdy 2000; van Pelt *et al.,* 2004), and other factors, i.e. culture density (Wagenaar *et al.,* 2006).

Different spatio-temporally recurring patterns occur in both stimulus-induced (Ferrández *et al*., 2013) and spontaneous activity. They are usually dynamic over time (i.e. the spatial location of active sites may change), thereby having different, yet characteristic spatio-temporal shapes (Shahaf and Marom, 2001; van Pelt *et al.,* 2005; Sun *et al.,* 2010; DeMarse *et al*., 2001; Pasquale *et al.,* 2008; Pasquale *et al.,* 2010; Ruaro *et al.,* 2005; Nadasdy 2000; Nomura *et al.,* 2009).

Without any external stimulus, cultured neurons show significant changes in their spontaneous neural activity at different stages toward maturity. Moreover, network activity fluctuations at later stages may be a consequence of repetitive internal stimuli that revive prior network activity and are thought to alter network connectivity to compensate for the lack of external stimuli. Such self-organized events based on spontaneous neural activity were previously reported at different culture ages (Rolston *et al.,* 2007; Pasquale *et al.,* 2010; Sun *et al.,* 2010).

Both, *in vivo and vitro*, stimuli may trigger bursting (Krahe 2004; Akerberg 2011) and PS activity. The early visual system is a prominent example. PS activity in retinal ganglion cells is driving suprathreshold responses at postsynaptic targets in the lateral geniculate nucleus (Usrey *et al.,* 1998; Sincich *et al.,* 2007; Weyand, 2007). PS enhancement contributes to preserving the information of a visual stimulus from one processing stage to the next (Rathbun *et al.,* 2010; Sincich *et al.,* 2009; Uglesich *et al.,* 2009). It has been shown that the second spike in a pair evoked a postsynaptic potential with maximum efficacy for inter-spike intervals (ISIs) in the range of 2 - 5 ms. Efficacy rapidly decreased to zero for ISIs larger than 40 ms (Usrey *et al.,* 1998; Sincich *et al.,* 2007).

However, little is known on the evolution and role of PS activity in neural cultures derived from dissociated brain tissue, on its relationship to bursting activity and on its participation in the organization of functional and effective network connectivity. To address these questions, we defined activity consisting of two spikes being separated by an interval of up to 5 ms followed by an inter-paired-spike interval (IPSI) larger than 40 ms as PS activity (Methods 2.3). We then analyzed 58 streams of continuously extracellularly recorded spontaneous neural activity in random networks for PS occurrence and for the spatio-temporal evolution of PS activity patterns over several weeks. In this context, we wondered whether any PS-induced effect was locally confined or led to changes on network level. We finally investigated the robustness of PS activity and its independence in driving spontaneous neural activity, thereby affecting functional and effective connectivity.

## 2. Material and methods

### 2.1 Continuous 59-channel MEA electrophysiology and spike train assembly

The data for this analysis was provided by a recently developed cell culture perfusion system that allowed us to continuously track both network activity and morphology on the lab bench at ambient $CO_2$ levels under rather well controlled environmental conditions (i.e. pH, temperature and osmolality). Technological and procedural details will appear in a dedicated article (Saalfrank *et al.*, submitted). With this setup, the activity evolution in a hippocampal and a cortical network on MEAs was continuously recorded over 30 and 53 days *in vitro* (DIV), respectively. These datasets were analyzed for PS activity. To reduce data file size, only upward (positive) and downward (negative) spike cutouts from 57 (cortical) and 58 (hippocampal) out of 59 recording electrodes were stored in 5 min packets. They consisted of 5 ms pre-spike and 5 ms post-spike fragments after first threshold crossing at ± 5.5 SD with respect to peak-to-peak noise (Suppl. Fig. 1(A)). Only timestamps from downward threshold-crossings were extracted using NeuroExplorer (Nex Technologies). After removing simultaneous timestamps that occurred on all channels due to electrical or handling artefacts, subsequent 5 min datasets comprised of ≤ 58 timestamp streams were bundled in 12 hour timestamp packets for further analysis in Matlab (MathWorks). In the following, these half-day packets will be called trials of duration $T_{Trial} \leq$ 12 h. On some days, trials encompassed less than 12 h due to temporary interruptions for system reconfiguration, maintenance work or power failure. Our recording sessions consisted of 65 trials (32.5 DIVs) for the hippocampal culture and 106 trials (53 DIVs) for the cortical culture.

### 2.2 Local and network firing and burst rates

Firstly, we quantified the local firing rates (LFRs) at individual sites as the number of recorded spikes divided by $T_{Trial}$ for each local spike train (LST) (Suppl. Fig. 1(B)). At network level, we pooled all spikes from all 57 and 58 sites, respectively, for each trial into a single network spike train (NST) by sorting them in a time-ascending order. The NST represented the MEA-wide activity for each trial. The network firing rate (NFR) was then quantified as the total number of spikes in an NST divided by $T_{Trial}$ (Suppl. Fig. 1(C)).

To further investigate activity dynamics, we used the burst rate (BR), a well-known parameter for characterizing synchronous network activity. We scanned all LSTs at the 57 (cortical) and 58 (hippocampal) individual sites for each trial and defined bursting activity as events with more than 10 subsequent spikes being individually separated by an ISI of less than 100 ms, followed by an interburst interval (IBI) larger than 200 ms (Wagenaar *et al.,* 2005) (Suppl. Fig. 1(C)). The local burst rate (LBR) at individual sites was calculated by dividing the number of bursts by $T_{Trial}$. Equally, the network burst rate (NBR) was obtained by scanning the NST for bursts using above mentioned criterion and dividing the number of bursts by $T_{Trial}$.

### 2.3 Paired spiking activity

Based on the previously described findings by Ursey (Usrey *et al.,* 1998) and Sincich (Sincich *et al.,* 2007), we investigated the occurrence and effect of paired spiking on neural activity in network cultures. As sketched in Suppl. Fig. 1(D), we defined PSs at individual sites as the neural activity consisting of two spikes recorded from the same electrode separated by an interval of up to 5 ms, followed by an IPSI larger than 40 ms (in order to assure that a second spike in a PS does not influence a first spike in a second PS for two consecutive PSs).

Equally, we scanned all the NSTs using the above mentioned algorithm to quantify the PS activity at network level for each trial. In this case, the two spikes separated by an interval of up to 5 ms and followed by an IPSI larger than 40 ms did not necessarily have to be recorded from the same electrode.

To describe the PS activity dynamics for each trial, we calculated the number of active sites with PSs ($NAS_{PS}$) as being the number of sites with at least two PS repetitions during $T_{Trial}$. Equally, we calculated the $NAS_B$ for bursts as the number of sites with more than one burst per trial.

To check if PS activity forms robust spatiotemporal patterns, we firstly calculated the IPSI as being the difference between two consecutive PSs at both network level and individual sites. From the IPSI histograms for each trial we extracted the highest number of IPSI repetitions and the most frequently encountered IPSI value ($IPSI_{mfo}$) at individual sites.

Moreover, to confirm that PS activity was neither governed strictly by firing rates (and thus represented an intrinsic neural response property) nor, at network level, by chance as a procedural result of projecting spiking activity from individual sites onto a single NST timeline, we generated Poisson-like network spike trains for comparison. In these, the firing probability was distributed according to a homogenous Poisson process without refractory period. If PS activity were strictly governed by firing rate, Poisson-like spike trains with the same firing rate as the recorded NSTs would give a similar PS distribution. Additionally, we shuffled the NSTs 100 times for each trial and quantified PS activity to check the degree of randomness of NST PS activity. NST spike times were randomly rearranged with the randperm (Matlab, MathWorks) function. 100 repetitions were chosen to warrant statistical significance at acceptable computational costs.

To generate above mentioned Poisson-like spike trains, we used Matlab user-written routines. The Poisson distribution P represents the probability that a homogenous Poisson process generates n spikes in a period of trial duration $T_{Trial}$:

$$P(n) = \frac{(rT_{Trial})^n}{n!}\exp(-rT_{Trial}) \qquad \text{Eq. 1a}$$

where r is the spike count rate defined as the total number of spikes divided by $T_{Trial}$ for each LST and NST.

Timestamps were generated by the following interspike interval formula:

$$t_{i+1} - t_i = -\frac{1}{r}\ln(rand) \qquad \text{Eq. 1b}$$

where rand is a random number uniformly distributed over the open interval (0 : 1); $t_i$ represent the spike timestamps for i = 1, 2,..., n spikes (Martiniuc and Knoll, 2012).

### 2.4 Post-stimulus time histogram and time-varying firing rates at network level

To investigate the hypothesis that PS activity might replace external stimuli sources, we considered each PS a stimulus-resembling event for the network. Thus, for this particular investigation at network level, we considered PS onset (first spike) as the beginning of a stimulus (t = 0 s) with a duration of $T_{PS}$ = 2 seconds, which is close to the shortest IPSI duration found for each trial.

To calculate the post-stimulus time histograms (PSTHs), the timestamps of PS-elicited spikes during a $T_{PS}$ were aligned relative to t = 0 s for each period $T_{PS}$. n reflects the number of PS stimuli at network level in a trial. $n_{min}$ was found to be 200. We divided the stimulus period $T_{PS}$ into N bins of duration $\Delta t$ = 5 ms and counted the number of spikes $k_i$ from all n sequences that fall into bin i. After averaging for the n stimulus repetitions and dividing by bin duration Δt, we obtained the time-varying firing rates r(t) with respect to stimulus (PS) onset.

## 2.5　Information content per spike

In order to evaluate the information about the stimulus (PS) carried by individual spikes following a PS within $T_{PS}$, we used the above calculated time varying firing rates r(t) and computed the entropy estimates (H) as follows (Strong *et al.*, 1998; Brenner et. *al.*, 2000; Sincich *et al.*, 2009):

$$H = \frac{1}{T_{PS}} \int_0^{T_{PS}} \frac{r(t)}{<r>} \log_2 \frac{r(t)}{<r>} dt \qquad \text{Eq. 2}$$

where $T_{PS} = 2$ s represents the above mentioned duration of a stimulus and $<r>$ the average firing rate; also in this case, the bin size was Δt = 5 ms.

For each particular trial, we obtained the average estimate for the information content per spike by averaging the estimated entropy by the number of stimulus repetitions n:

$$\langle H \rangle = \frac{1}{n} \sum_{i=1}^{n} H_i \qquad \text{Eq. 3}$$

To account for limited dataset size and to correct the resulting bias, the information content per spike was estimated as a function of bin size Δt. We performed a linear fit to these data to extract the intercept corresponding to the limit when Δt approaches zero. We used the shuffle verification method to check for the robustness of information content per spike. Briefly, we randomly rearranged each NST as described in section 2.3 and computed estimates of the information content per spike as mentioned above (Eq. 2). For each of the shuffled NSTs, we repeated this procedure 100 times and obtained a standard deviation of estimated information content that was smaller than one standard deviation of the fitting intercept obtained from above mentioned linear fit.

This measure of information content does not make any assumption about the stimulus features; it only reveals the information content carried by individual spikes.

## 2.6　Conditional firing probability

The highly variable spontaneous spiking activity of cultured neurons features robust patterns (i.e. bursting activity), which might participate in the establishment of functional connections between different sites within the culture. To investigate whether PS activity plays a role in shaping the dynamic interconnectivity map at different developmental stages, we adopted a variant of a cross-correlation algorithm initially introduced as conditional firing probability (CFP; le Feber *et al.,* 2007). This method was widely used in the investigation of activity relationships between different electrodes to reveal the formation and strength evolution of functional connectivity within *in vitro* networks (Zullo L *et al.,* 2012; Chiappalone *et al.,* 2007; Garofalo *et al.,* 2009). Here, we used CFP to reveal any PS-related cross-correlations between different sites within the network. That is, at each electrode i (i = 1 : 57 or 58, respectively) considered as the reference, we selected the second spike in each PS as a reference with new relative time $t_i = 0$. We then calculated the CFP as the probability of spike occurrences at any of the other 56 or 57 recording electrodes j (j = 1: 56 or 57, respectively) within the time interval $T_{CFP}$ [$t_i : t_i + 500$ ms] divided by the total number of second reference spikes of a PS at reference electrode i over the entire trial duration of $T_{Trial} = 12$ h. The spikes found at electrode j during $T_{CFP}$ were aligned relative to each $t_i$ and binned with a bin size of Δt = 1 ms. If any of the resulting 57*57 or 58*58 CFP(i,j) distribution curves showed a clear peak, we considered electrode j being correlated to the PS activity on reference electrode i. The peak amplitude was a measure of correlation strength. Its timestamp reflected the PS-related synchronization delay between the two neurons.

Additionally, two boundary conditions were chosen as restrictive validity criteria: a CFP(i,j) was rejected if the width at 80% of the peak value was shorter than 5 ms (five bin sizes; to avoid false correlations caused by outliers) and for synchronization delays larger than 250 ms (to avoid curves that decreased to zero beyond the 500 ms window).

## 2.7  Transfer entropy

Above mentioned concepts (i.e. information content per spike and CFP) quantify statistical dependences of observed variables (i.e. recorded spike trains), thereby describing functional connectivity maps which do not allow us to investigate the direction of information flow (i.e. the causality) between the recorded units. By definition (Wienner, 1956), an effective connectivity between two neurons exists when knowledge about the past of one neuron predicts the future activity of its counterpart better than the prediction based on the past activity of the receiver (neuron) alone. This effective connectivity is quantified by an information–theoretic measure called transfer entropy (TE) that was introduced by Schreiber (Schreiber, 2000). TE is an asymmetric measure of interactions between two coupled neurons which reveals effective connectivities and indicates the direction of information flow between recorded units. It permits to predict the spiking activity of a post-synaptic neuron by taking past spiking activity of its pre-synaptic partner into account. In our work, TE is positive and thus the information is directed from a sender unit A to a receiver unit B (i.e. there is an effective connectivity from unit A to unit B) only when the information about the spiking activity recorded at unit A improves the prediction of the spiking activity in the future of unit B better than any prediction derived from past spiking activity recorded at the unit B alone. Thus, to identify and assess effective connectivity within the neural network from recorded spiking activity, we used a recently introduced toolbox for calculating the TE (Ito et al., 2011) derived from the original definition given by Schreiber (Schreiber, 2000) as follows:

$$TE_{A \to B} = \sum p(B_{t+1}, B_t^k, A_t^l) \log_2 \frac{p(B_{t+1}|B_t^k, A_t^l)}{p(B_{t+1}|B_t^k)}$$

Eq. 4

A complete description of the TE toolbox algorithm can be found in Ito et al., 2011. Briefly, p describes a probability, $A_t$ depicts whether at time t a spike at unit A was recorded (and thus $A_t = 1$) or not ($A_t = 0$). Similarly, $B_t$ and $B_{t+1}$ describe the status of unit B at times t and t+1. Conditional probabilities of observing the particular status of units A and B are marked by vertical bars while the sum is over all possible combinations $B_{t+1}$, $B_t^k$ and $A_t^l$, where parameters k and l express the number of time bins in the past that allow us to take the time delay and the message length into account when calculating TE. For biophysical reasonability, we chose k = 1:30 ms and l =1:250 ms.

In this general framework, we exclusively considered PS activity at unit A (the sender) while unit B (the receiver) encompassed the entire recorded spiking activity. In this way, we could estimate whether PS activity at unit A was involved in information transfer toward unit B. Furthermore, we exemplarily chose the eight closest recording units as depicted in Figure 6A to check if PS was involved in information transfer and thus in establishing effective connectivity between these eight closest neighbors within the network. Thus, each of the closest eight units was scanned for PS activity and considered as the sender with respect to the entire spiking activity of the remaining seven closest neighbors. This resulted in a TE map, which depicts the PS information transfer dynamics of each of the selected senders (eight units A) toward the selected receivers (seven units B). For computational reasons, we split each trial duration T into 30 minutes subsets of recorded data.

We applied the same algorithm at network level to investigate the effect of information transfer from the above mentioned eight units A toward the rest of the network. In this case, we calculated the PS-related TE for each of the eight selected channels with respect to the entire NST as the only receiver unit B.

Further on, we calculated the differences between the resulting TEs:

$$\Delta TE = TE_{A \to B} - TE_{B \to A} \qquad \text{Eq. 5}$$

When ΔTE is positive, the information transfer is directed from A to B; in the opposite case, the information flows from B to A.

## 3. Results

### 3.1 Evolution of firing and burst rates through different developmental stages

Taking advantage of the uninterrupted extracellular recording technology for cultured neurons based on 59-channel microelectrode arrays (MEA), we analyzed the day to day evolution of spontaneous neural activity at both individual sites and network level. We extracellularly recorded activity from two different networks cultured under similar conditions. Quasi-continuous datasets from 7 DIV (first extracellularly recorded spikes emerged from the 5 µV noise floor and crossed the -5.5 SD of the peak-to-peak noise spike detection threshold) to 39 DIV for the hippocampal culture and from 24 DIV to 77 DIV for the cortical culture were analyzed.

For the hippocampal culture, the evolution of spiking activity at network level could be clearly divided into three periods (Figure 1A – blue bars) as follows: 7 DIV to 14 DIV as the first period (C1-1), 15 DIV to 26 DIV as the second period (C1-2) and 27 DIV to 39 DIV as the last period (C1-3). An obvious finding was a significantly increasing network firing rate (NFR – Methods 2.2) ($p < 0.05$, t-test) from one period to the next, starting with a mean firing rate of 1.2 spikes/s in the first period to 12.7 spikes/s in the last as indicated by the yellow mean values in Figure 1A. This increasing spontaneous spiking activity in maturing cultures is in accordance with previously reported results (van Pelt *et al.,* 2005). Significant changes have also been found at individual sites where the number of active sites during the same developmental periods increased as the culture grew toward maturity. In contrast, we found fluctuating neural activity periods containing or terminating with high network firing rates (Figure 1A, red bars) for the more mature cortical culture. We could distinguish six time periods marked by an increasing period followed by a decreasing trend. The individual periods lasted from 24 DIV to 37 DIV (C2-1) with a mean of 15±8 spikes/s, from 38 DIV to 45 DIV (C2-2) with a mean of 25±11 spikes/s, from 46 DIV to 50 DIV (C2-3) with a mean of 8.5±7 spikes/s, from 51 DIV to 65 DIV (C2-4) with a mean of 7±3.4 spikes/s, from 66 DIV to 71 DIV (C2-5) with a mean of 6±2 spikes/s and from 72 DIV to 77 DIV with a mean of 5.5±1.8 spikes/s for the last period (C2-6).

Firing rates are usually used to reveal characteristic communication mechanisms that are different for spontaneous and induced activity, respectively. In contrast, bursting activity plays a role in filtering spontaneous neural activity (van Pelt *et al.,* 2004; Wagenaar *et al.,* 2005). In our recordings, spiking activity tended to induce bursts of synchronized activity at different developmental stages.

Similarly to the network firing rate (NFR, Figure 1A, blue trace), the hippocampal culture showed a significantly ($p < 0.05$) increasing trend in bursting activity at network level (see Methods 2.2, Figure 1B, blue bars) over the three periods. A mean of 0.018 bursts/s during the first period increased to a mean of 0.22 bursts/s for the last period (Figure 1B, yellow dots). This developmental trend has already been reported in other network studies (van Pelt *et al.,* 2004). However, while the number of active bursting sites ($NAS_B$) increased from the first period to the second, it returned close to the value of the first period at the end of the recording. It dropped sharply during a power blackout (temperature dropped and stayed at room temperature for several hours), from which it recovered slowly to its previous value (see Figure 1D and corresponding yellow circles depicting means). In the second period, these extremely high $NAS_B$ are explained by massive neural avalanches that take place within the network and recruit neurons at most sites (ca. 82% of the 58 recording electrodes) for a short time. Figure 1C exemplarily shows such a network avalanche that arose at 18 DIV.

In the more mature cortical culture, we found less bursting activity at network level than in the hippocampal culture at earlier developmental stages, but with a high $NAS_B$. That is, a larger number of neurons contributed to the network bursting, but with a lower number of bursts/s, which did not lead to a comparable increase in the NBR. In addition, the NBR, $NAS_B$ and NFR of the cortical culture oscillated within each of the six periods, as indicated in Figure 1A, B and D (red bars).

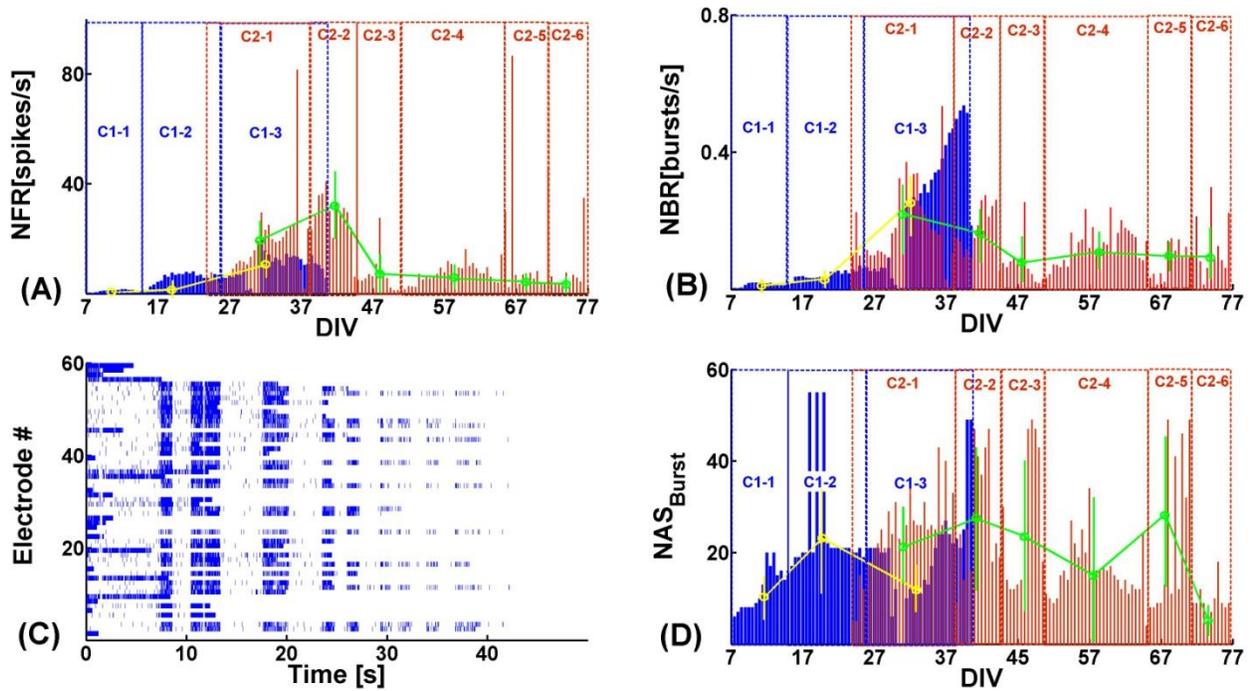

Figure 1 Evolution of network firing rate (NFR) (**A**) and network burst rate (**B**) over time for the first hippocampal (blue) and second cortical (red) culture. Three (hippocampal, blue rectangular delimiters, C1-1 to C1-3) and six (cortical, red rectangular delimiters, C2-1 to C2-6) recording periods were distinguished by significant changes in their NFRs. For each period, the mean NFRs and their SDs are displayed as circles with error bars. (**C**) Example of a network avalanche at 18 DIV.(**D**) Number of active sites ($NAS_B$) with respect to bursting activity in the hippocampal (blue) and cortical (red) culture.

### 3.2 Evolution of paired spiking activity at individual sites and at network level

Activity patterns consisting of PSs separated by ISIs of up to 5 ms were rarely encountered in the young hippocampal culture. Instead, random isolated spiking rather than synchronized rapid firing dominated neural activity as reported before (van Pelt *et al.*, 2005). While spike pairing was very low at early DIVs, it consistently increased after 3-4 weeks *in vitro* (WIV). This trend was robust not only at individual sites (Figure 2C, blue bars), but also at network level (Figure 2D, blue bars). The mean of PS occurrences at individual sites significantly ($p < 0.05$) increased about thirtyfold from 193 for the first period to 5741 for the last period (Figure 2C, yellow circles indicating mean highest number of PS at individual sites). At network level, the mean number of PSs in the first period was 486 and increased tenfold to 4911 in the last period (Figure 2D, yellow circles). In contrast, PS activity in the more mature cortical culture did not grow monotonically, but fluctuated rather synchronously with both the firing and burst rates at network level (Figure 2C, D, red bars; Figure 1A, B).

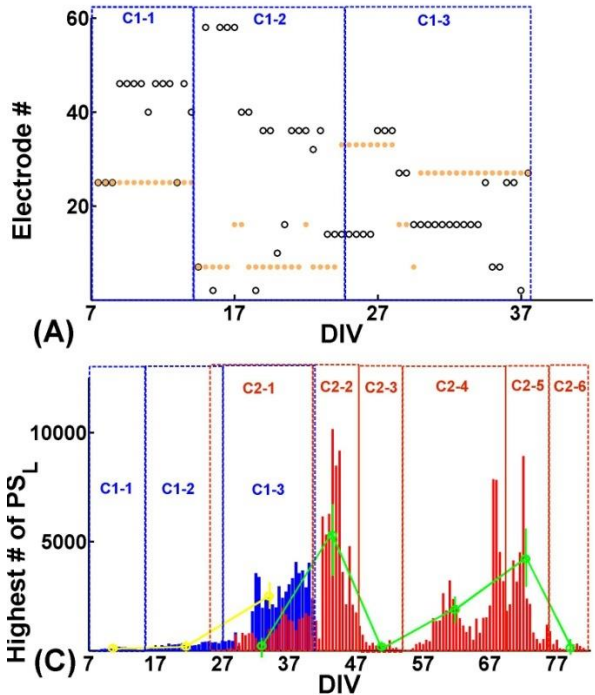
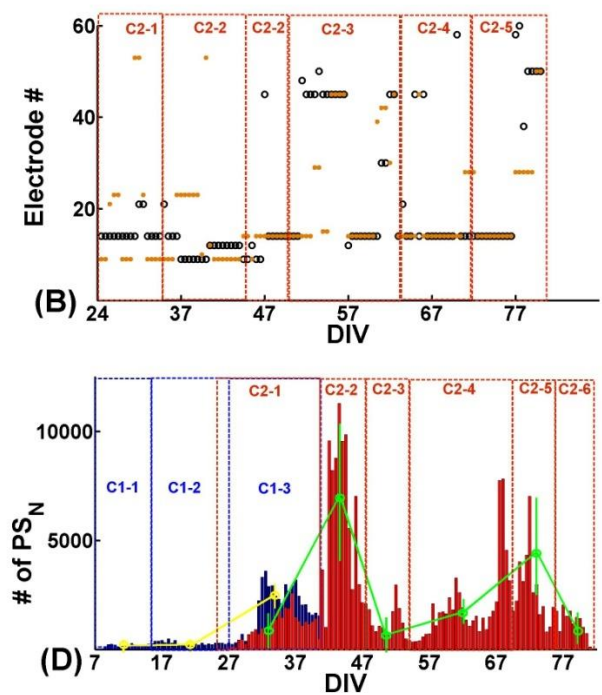

Figure 2 Highest PS activity (black circles) and highest bursting activity (orange circles) in the hippocampal (**A**) and cortical (**B**) culture may not necessarily be recorded from the same electrode in subsequent trials. (**C**) Number of occurrences of the most frequently, locally occurring $PS_L$ in the hippocampal (C1, blue bars) and cortical culture (C2, red bars) at the individual recording sites denoted on the y-axes in (A) and (B). (**D**) Total number of $PS_N$ occurrences at network level for the hippocampal (C1, blue bars) and the cortical culture (C2, red bars). Circles and error bars in (C) and (D) display mean PS values and their SDs (yellow: hippocampal; green: cortical).

Interestingly, the IPSI (see Methods2.3), a parameter which quantifies the temporal gap between two consecutive PS, exemplarily suggests that PS activity becomes robust as the culture ages. For the hippocampal culture, the duration of the most frequently occurring (mfo) $IPSIs_{mfo}$ at both individual sites ($\leq 90$ s, index 'L') and network level ($\leq 48$ s, index 'N') was very long and fluctuated highly with a low number of repetitions in the first three WIVs (Figure 3A, individual sites; Figure 3B, network level, blue bars). The non-uniform temporal distribution of PS activity during this developmental period denotes that PS was not yet robust. Remarkably, $PS_N$ activity at network level shows higher than average values exactly at those DIVs with strong activity avalanches (as exemplified in Figure 1C at 18 DIV). This suggests that such extremely high neuronal activity at individual sites (as revealed by the $NAS_{PS}$ in Figure 4A, blue bars) accounts for the elevated $PS_N$ activity at network level, while firing rates kept a uniformly increasing trend at those DIVs (Figure 1A). After three WIVs, PS activity dramatically increased (Figure 2C, individual sites; Figure 2D, network level, blue bars), while the duration of the most frequently occurring $IPSIs_{mfo}$ decreased and robustly settled at 2-3 s until the end of the recording session (Figure 3A, individual sites; Figure 3B, network level, blue bars). Interestingly, the duration of the $IPSIs_{mfo}$ at network level (Figure 3B) stabilized earlier than at individual sites (Figure 3A). Over the same period, the number of the $IPSIs_{mfo}$ increased consistently both at individual sites and network level (Figure 3C and D, blue bars). These three trends (increasing overall number of PS, decreasing duration of $IPSI_{mfo}$, increasing number of the most frequently occurring $IPSI_{mfo}$) suggest that PS activity develops homogeneously and consistently throughout the network to result in robust PS activity patterns with an increasing number of occurrences with rather constant ISIs and IPSIs, especially after three WIVs.

For the more mature cortical culture, we observed the same inverse correlation between duration and number of the $IPSIs_{mfo}$ (see Figure 3A and D, red bars), however, oscillating over time. There were recurring periods with large IPSI values and low numbers of IPSI repetitions followed by periods with lower IPSI values but high repetition frequencies.

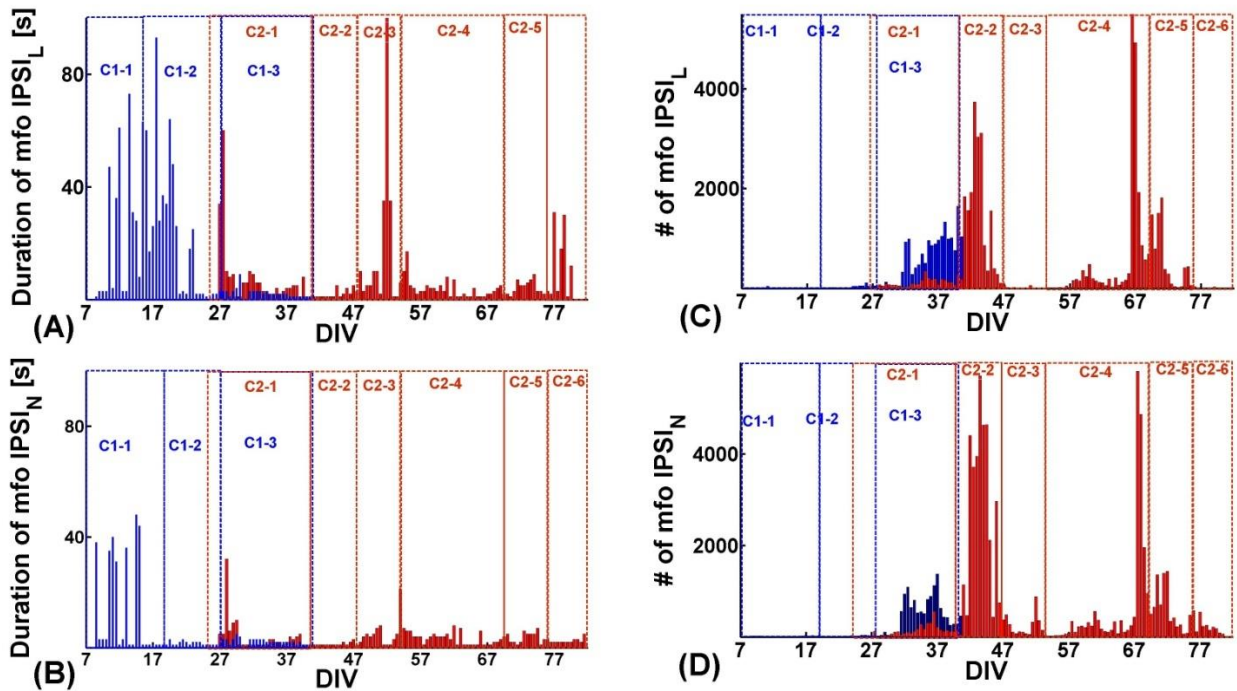

Figure 3 Duration of the most frequently occurring (mfo) $IPSI_{mfo}$ at individual sites (**A**) and at network level (**B**) for the hippocampal culture (C1, blue bars) and the cortical culture (C2, red bars) and its evolution from one period to the next (C1-1 to C1-3 and C2-1 to C2-6, respectively). Number of the $IPSI_{mfo}$ at individual sites (**C**) and at network level (**D**) (with the same color and period coding as in (**A**) and (**B**)).

### 3.3 PS activity versus burst activity

In order to check if PS activity is one of the driving forces for the self-organization of functional network connectivity, we investigated the relationship between bursting and PS activity. For the hippocampal culture, on average only 11% of bursts contained PS at early DIVs. Their number slightly increased to 16% in the last recording period. Except for four trials (6.25% of total trials), PS activity could be found in less than 40% of the bursts (see Figure 4B, blue bars). The same trend was observed for the cortical culture where, except for seven trials (3.18% of total trials), the percentage of bursts that contained PS remained below 50% (see Figure 4B, red bars).

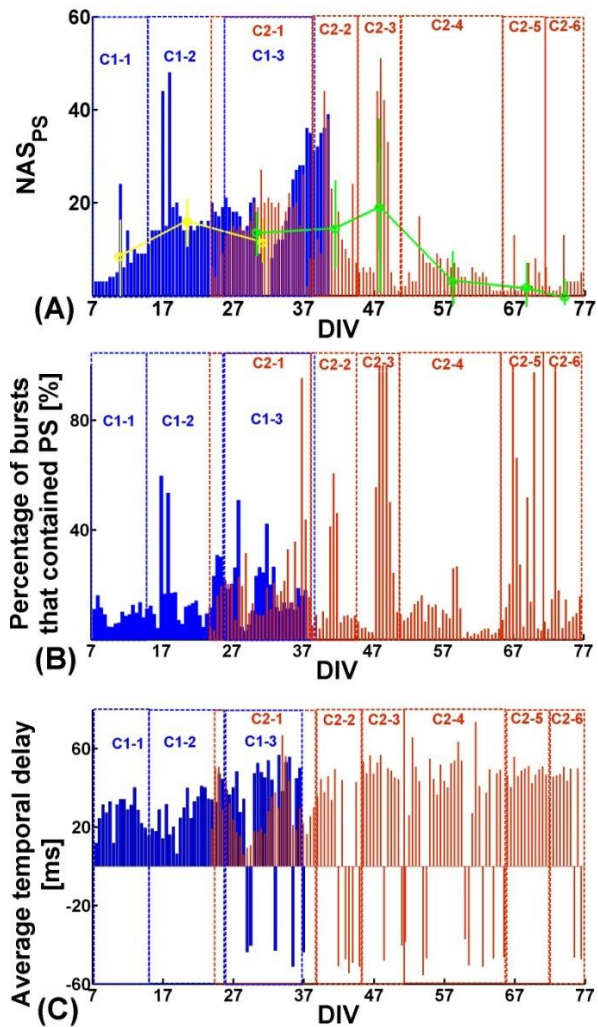

Figure 4 (**A**) NAS for PS activity for the hippocampal (blue) and the cortical (red) culture. (**B**) Percentage of bursts that contained PS at individual site level. (**C**) Average temporal delay in ms with which a burst followed a PS at individual sites.

For both the hippocampal and cortical network, PS and bursting activity were present at most sites (Figure 5A). However, in most cases, highest PS activity was recorded from different electrodes than those that recorded the highest bursting activity, as pointed out by the black circles (PS) and orange circles (burst) in Figure 2A (hippocampal) and Figure 2B (cortical). For the hippocampal culture, highest PS and bursting activity were spatially collocated in only six trials (9.3% of total trials). In contrast, the electrodes with highest PS and bursting activity coincided in 42 trials (38.18 % of total trials) for the cortical culture, 41 of them occurred after 44 DIV. In both cases, the network location with dominant PS and bursting activity could change over the days. Over the course of the entire recording, highest PS activity was detected on 12 different electrodes (21%) for the hippocampal culture and on 11 electrodes (19%) for the cortical culture. Highest bursting activity could be associated with just five electrodes (8.3%) in the hippocampal culture (Figure 2A) while it occurred on 12 electrodes (20%) in the cortical culture (Figure 2B).

Importantly, in almost 92% of the recording trials PS activity preceded bursting activity (Figure 4C, blue bars) in the younger hippocampal culture and in 84% of the recording trials for the more mature cortical culture (Figure 4C, red bars). Additionally, the average temporal delay between a PS and a burst mostly remained below 50 ms. This suggests that PS presumably initiated bursting activity. Interestingly, this PS-burst coupling occurred on the same electrode of the cortical culture in 34 instances, suggesting robustness of PS-dominant sites. This electrode also recorded highest PS activity in almost 50% of the trials.

We further investigated the stability of the spatio-temporal distribution of PS patterns. The middle insets in Figure 5 (i1 - i6) exemplarily show color-coded PS spiking activity maps at individual sites in the hippocampal culture for six trials at different developmental stages. Robust spatial patterns of PS activity were found over the entire recording period; two examples are highlighted by blue circles.

A total of 12 sites with stable PS patterns could be identified in more than 50% of the 60 trials. Among these sites, eight lasted longer than 73% of the total recording period. They formed robust, long-lasting patterns that presumably recruited new neighboring sites in different trials. Furthermore, seven of these sites also formed robust burst patterns as marked by yellow circles in Figure 5A.

Figure 5B exemplarily shows PS activity patterns that were formed during the first period (yellow circles) and lasted until the end of the recording session, patterns that were newly formed during the second period and lasted until the end (black circles) and new neighboring sites that emerged only during the third recording period (white circles). Thus, 10 sites formed patterns that lasted for more than four trials during the first period, 22 sites for the second period and 34 sites for the last period. That is, for each new period, up to 12 neighboring sites were recruited in generating PS activity, thereby increasing the degree of PS pattern complexity as the network entered later developmental stages.

### 3.4 Information content per spike and CFP analysis

Next we asked to what degree spontaneous *in vitro* PS activity preserves its role encountered *in vivo* and thus participates in the formation of functional connectivity and in information processing within the cultured neural network at different developmental stages. With this motivation in mind, we considered PS activity as an internal stimulus and thus calculated the PS-related CFP (see Methods 2.6) and PS-related information content (see Methods 2.5) in the hippocampal network both at individual sites and at network level.

Because 12 sites presented robust, long-lasting PS activity for more than 50% of the total trials (Figure 5A), we exemplarily calculated the PS-related correlation between eight closest neighbors out of these 12 sites as indicated in Figure 6A (red ellipses).

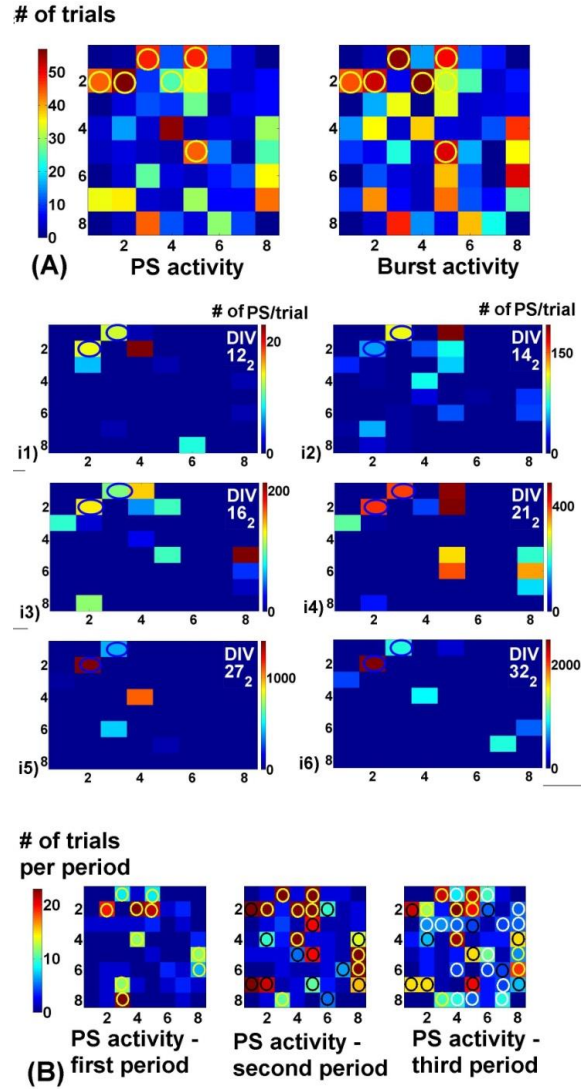

Figure 5 (**A**) Spatial PS and burst pattern distribution with respect to the 8 x 8 electrode matrix for the hippocampal culture. Yellow circles mark the seven electrodes from which both PS and bursting activity could be recorded according to the criterion described in Methods 2.2 and 2.3. One trial represents half a DIV (Methods 2.1). Marked electrodes recorded activity in the majority of the trials, though not necessarily consecutively. (**B**) Evolution of PS pattern complexity in the hippocampal culture from one period to the next: the NAS with PS patterns increased during development. In most cases, new PS emerged on electrodes adjacent to those with previous PS activity. For the second, more mature cortical culture, 54 sites (93%) participated in PS activity that lasted for at least two trials, while in more than 50% of the trials the number of sites decreased to four. This suggests that at later developmental stages the role of dominant sites gains importance.

We used CFP to construct the inter-connectivity maps for both the interconnectivity strength (Figure 6B) and the temporal delay (Figure 6C). For each trial and site i (i = 1:8), we quantified the PS-related CFP (i,j) for all possible pairs (j = 1:7). Interestingly, we found no connectivity during the first 11 DIV, coinciding with the period in which PS activity was not robust yet (i.e. with large IPSIs and few repetitions). In contrast, from 17 DIV onward the number of connections significantly increased (p < 0.01) and remained high until 27 DIV. This corresponds to the second period, where PS activity gained robustness. The highest number of connections was found in this period (Figure 6D), which decreased thereafter. Moreover, the temporal delay of the correlations between formed pairs increased until 18 DIV with a mean of up to 45.2 ms (±20 ms) and consistently decreased thereafter with a mean of 16.2 ms (±10 ms) (Figure 6E and F). Connectivity strength stayed rather constant over several DIVs with almost identical means of around 0.07 (±0.03) for all three recording periods (Figure 6E).

Next we looked at the PS-related information content per spike (Eq. 2, Methods 2.5) for each selected channel pair (i,j) of the eight interconnected sites. The resulting information content map is presented in Figure 8(A). Information content per spike was highest during the second period, thereby correlating with the highest numbers of connections between these eight most active channels (Figure 8B compared with Figure 6B and D). This trend is also reflected by the mean values (Figure 8B). The initial increase in information content per spike from 0.8±0.2 bits/spike during the first period to 2.2±0.3 bits/spike in the second period is followed by a decrease to 1.2±0.6 bits/spike during the last period. While this statistical measure associates the information content per spike with PS, it does not make any statement on how the content is actually carried by the PS and on whether the PS is the only information carrying mechanism.

The three observations, i) the decreasing trend in temporal delay between PS-induced correlated activity, ii) an almost constant interconnectivity strength at later developmental stages and iii) the formation of robust spatiotemporal PS activity patterns as the culture matured may indicate that PS activity participates in the development and stabilization of functional connections at individual sites. To check for the robustness of the PS-related connectivity map, we constructed artificial Poisson-like spike trains according to Eq. 1 (Methods 2.3) for these eight electrodes (Figure 6A, red circles) for all trials.

Next we investigated whether the PS-related connectivity trend at individual sites is also found at network level for the different developmental stages. Firstly, we checked the robustness of each NST by asking whether PS activity at network level is a "by chance" result of mapping all spikes from individual sites onto a single timeline. We therefore shuffled all of the NSTs repeatedly for 100 times and quantified PS activity for each individual case. We then investigated whether the artificial spike trains that mimic the recorded spike trains (i.e. artificial spike trains have the same firing rates as the recorded ones) develop similar connectivity maps as the real spike trains. Robustly, we found no PS-related connectivity between the constructed spike trains for any of the trials in the artificial spike trains.

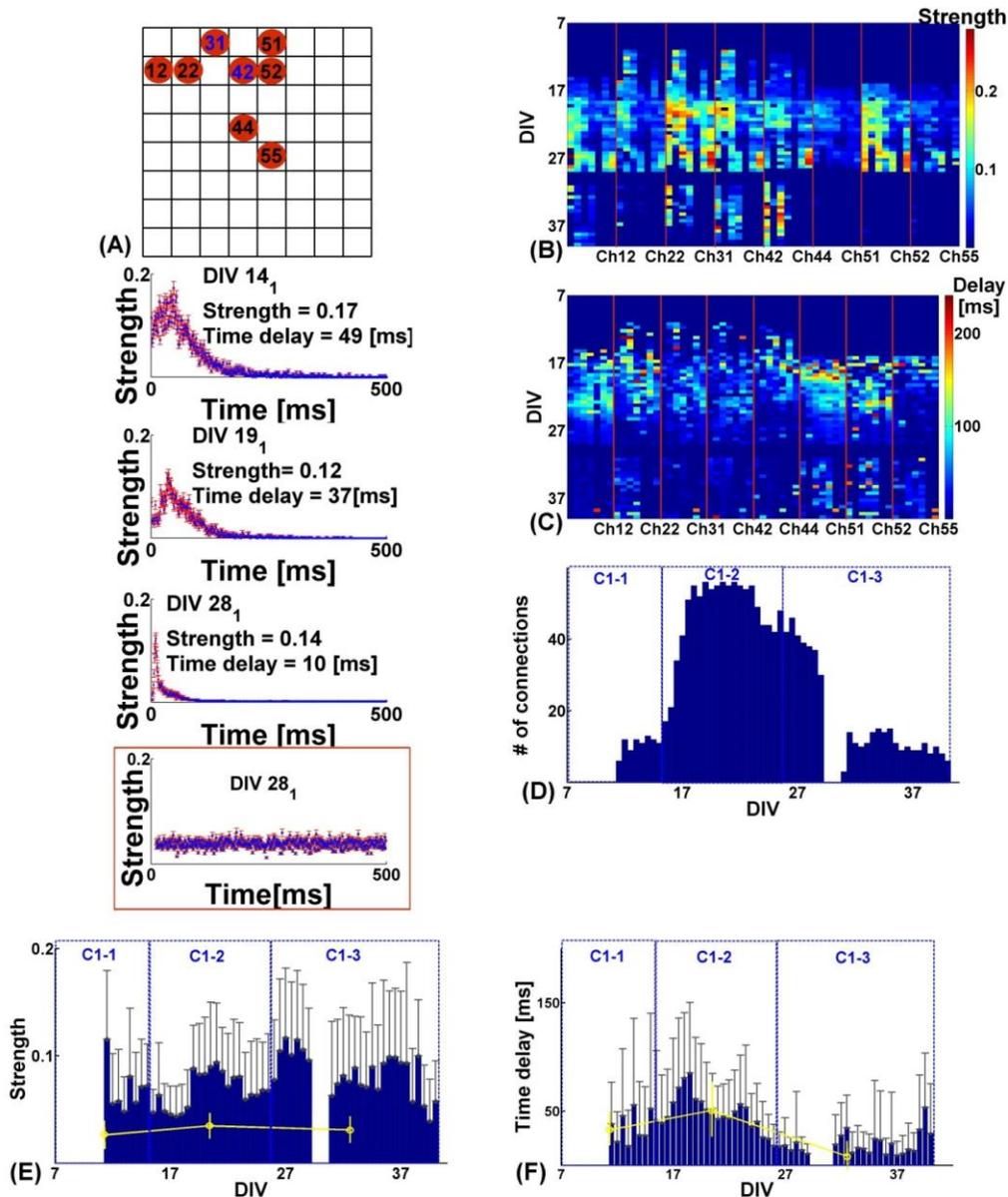

Figure 6 (**A**) Spatial arrangement of the eight closest, most PS-active sites in the hippocampal culture with respect to the 8x8 MEA matrix layout. The first number in a pair refers to the column, the second to the row. The insets under (A) exemplarily show PS-related CFP curves of reference channel 31 vs. channel 42 at three different DIVs (14, 19 and 28). The flat CFP curve framed by a red box illustrates the lack of PS-related connectivity for artificial spike trains that mimic channels 31 and 42. Connectivity evolution over time expressed in strength (**B**) and temporal response delay (**C**) between the exemplarily selected eight most active PS sites. Each pixel column represents one of the seven recording sites connected to the respective reference channel indicated on the x-axis and pointed out in (A). Sorting order is column, then row. The red vertical bars delimit the eight reference channel permutations. (**D**) Evolution of the number of connections between the selected channels. (**E**) Evolution of the average connectivity strength for the selected channels and their means for each period (yellow circles). (**F**) Evolution of the average connectivity time delays for the selected channels and their means for each period (yellow circles).

Insets of Figure 6A exemplarily show the connectivity between PS activity of channel 31 and spiking activity at channel 42 for three different DIVs. As mentioned before, while for the recorded spike trains the strength of the connectivity remained fairly constant, the time delay decreased during the last period (28 DIV). In contrast, the red box inset exemplarily shows no connectivity between the two artificial spike trains that mimic the same two electrode recordings over the same period.

We found the PS activity for each trial to be almost zero (Figure 7C, blue bars represent PS in recorded NSTs; inset with red bars represent PS in shuffled NSTs). To mimic the recorded NSTs, we further constructed artificial Poisson-like spike trains with similar firing rates as the NSTs (Eq.1, Methods 2.3). Also in this case, such artificial NSTs showed statistically significant (p<0.001) different PS activity as if it was strictly governed by firing rates (Figure 7C, green bars). These Poisson-like spike trains lack a spike history (i.e. without refractory period). Thus, very large NST firing rates and an exponential ISI distribution (Eq. 1b) favor short ISIs (i.e. up to five ms), which leads to an unrealistically high number of PS occurrences.

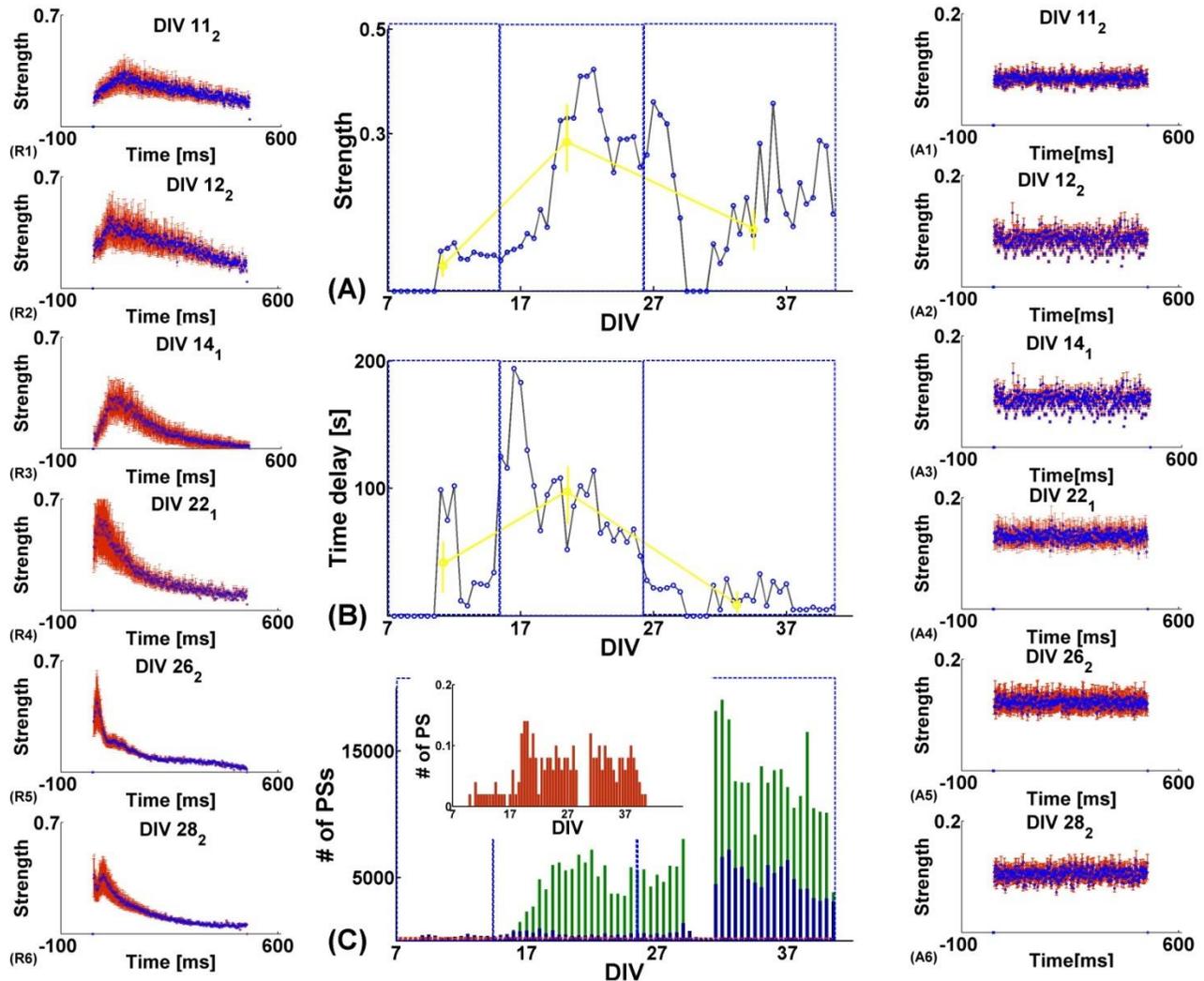

Figure 7 (**A**) Evolution of PS-related CFP strength for all hippocampal NSTs. (**B**) Evolution of PS related CFP time delay for all NSTs. (**C**) The number of PS for artificial spike trains at network level (green bars) and the number of PS at network level for recorded NSTs (blue bars). The inset shows a zoom onto the number of PSs for shuffled NSTs (red trace). The insets **R1 – R6** (left side column) display examples of PS-related CFPs at network level for different DIVs (11, 12, 14, 22, and 28). The insets **A1 – A6** (right side column) display examples of PS-related CFPs at network level for artificial spike trains for the same DIVs (11, 12, 14, 22, and 28).

Next, we checked for PS-correlated activity at network level by calculating the CFP (Methods 2.6) for every NST, this time with respect to developmental evolution of network-wide, PS-induced activity instead of local connectivity. As before, the second spike in a spike pair (that matched the PS criterion of ≤5 ms) of an NST served as the reference for calculating the CFP with respect to the following 500 ms of NST activity. This auto-correlation-like analysis provided information on the strength and time delay of PS-related spiking activity for individual NSTs. If repeated for all NSTs, the evolution of PS-correlated activity can be plotted for all trials (Figure 7A, B). Remarkably, we found that PS-related spiking activity started at 11 DIV for all of the trials. Its time delay increased until 18 DIV and significantly decreased during the last recording period, which is strikingly similar to the previously observed trend at individual sites as reported above.

The mean time delay (Figure 7B, yellow circles) increased during the second recording period to 74.5±30 ms and decreased in the third period to 12.4±10 ms while the mean correlation strength (Figure 7A, yellow circles) increased during the second period to 0.28±0.09 and decreased to 0.14±0.09 in the last recording period.

Furthermore, despite significantly larger PS activity found in artificial spike trains that were supposed to mimic NSTs, we found no correlated spiking activity in any of the trials (insets A1 – A6 to the right side of Figure 7). In contrast, insets R1 – R6 on the left side of Figure 7 exemplarily show the CFP of PS-related spiking activity for six of the NSTs indicating a decreasing time delay of the PS-correlated spiking activity.

Finally, after noticing that PS activity is involved in developing functional connections at individual sites and that spiking activity at network level is correlated with each second spike in a PS (with a decreasing time delay) for all NSTs, we checked whether PS activity may also be involved in information processing at network level. We calculated the PS-related information content per spike at network level for each NST considering PS activity as an internal stimulus (Eq. 2, Methods 2.5). Indeed, we found that in both cultures PS activity is involved in information processing at network level as well (Figure 8C). Moreover, in the hippocampal culture, the trend found at local sites was preserved at network level. For the first period, we found a mean of 3.4 bits/spike (SD = 1.3 bits/spike), which increased during the second period to 4.9 bits/spike (SD = 0.4 bits/spike) and decreased in the last period to 3.6 bits/spike (SD = 1.2 bits/spike). For the cortical culture, the information content increased and decreased for different periods from 4.4 bits/spike (SD = 1.8 bits/spike) up to a value of 6.5 bits/spike (SD = 0.8 bits/spike).

### 3.5 PS activity participates in controlling the direction of information flow within the coupled neuronal units

So far we have seen that PS activity is involved in establishing functional connectivity and carries information both at local side and at network level. We then asked whether PS is also controlling the direction of information flow within the cultured network. A TE analysis (Methods 2.7) may reveal how presynaptic PS activity predicts activity at its postsynaptic target or even at network level. We exemplarily choose the eight most closest recorded units mentioned above (as depicted in Figure 6) and calculated ΔTE (Eq. 5) for PS activity for each of the channels with respect to the entire spiking activity of the remaining seven channels for different time lags and message lengths (Ito et al., 2011). We found that PS activity is robustly involved in information transfer between cultured neurons.

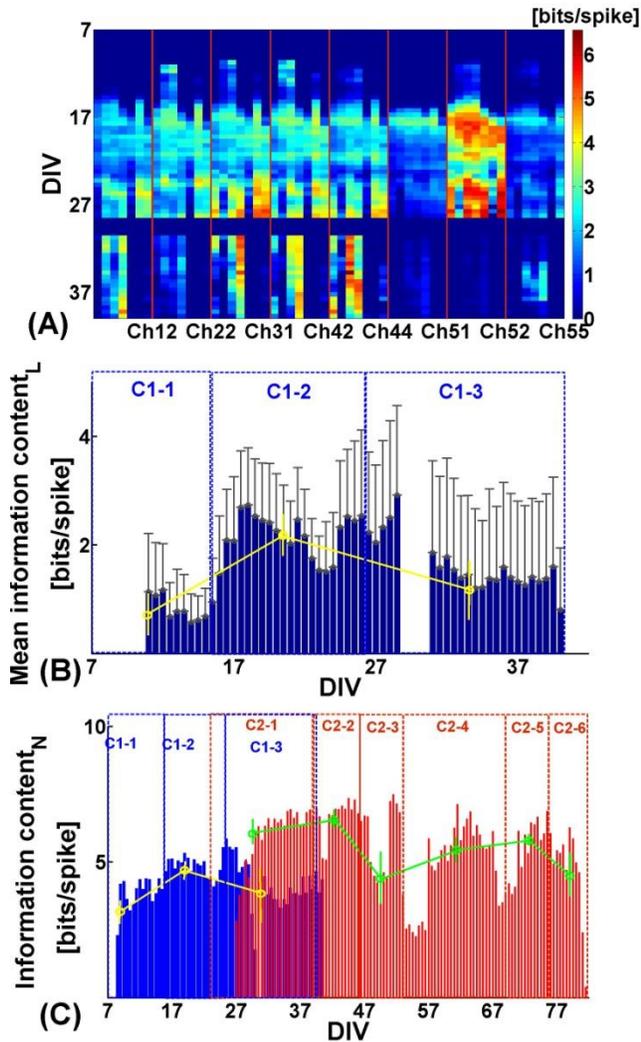

Figure 8 (**A**) Matrix of PS-related information content expressed in bits/spike between the exemplarily selected eight most active PS sites in the hippocampal culture (Figure 6). Each pixel column represents one of the seven recording sites being connected to the respective reference channel indicated on the x-axis. Sorting order is column, then row. The red vertical bars delimit the eight reference channel permutations. (**B**) Evolution of the PS-related mean information content in bits per spike (blue bars) for the selected channels and its respective average for a given period (yellow circles). (**C**) Evolution of information content for PS-related activity in both cultures at network level for each period. Blue bars represent the hippocampal culture with mean and standard deviations in yellow; red bars represent the cortical culture with mean and standard deviation in green.

Figure 9A shows the constructed ΔTE map, which indicates a very low effective connectivity between selected recording units at an early developmental stage (i.e. the first 17 DIVs) with a mean $\Delta TE = 0.2 \cdot 10^{-2}$ bits/s and $SD = 0.18 \cdot 10^{-2}$ bits/s. Additionally, for some channels, PS activity did not establish any effective connectivity during this period. Only three out of eight selected channels showed effective connectivities with their postsynaptic partners. Those disappeared and reappeared in different recording periods indicating that PS activity was not robust yet and consequently could not reliably predict or cause spiking activity at their targets. This situation changed dramatically from 17 DIV onward. ΔTE significantly increased ($p<0.001$) up to 29 DIV indicating that PS activity established robust and effective connectivities. Only on channel 51 PS activity established less effective connectivities during this developmental stage. For this developmental period we found a mean $\Delta TE = 1.13 \cdot 10^{-2}$ bits/s with a $SD = 0.14 \cdot 10^{-2}$ bits/s. From 30 DIV onward until the end of the recording period, ΔTE decreased with a mean of $0.4 \cdot 10^{-2}$ bits/s and a $SD = 0.13 \cdot 10^{-2}$ bits/s, while only four (50%) channels with PS activity had established effective connectivity with large ΔTEs. Remarkably, PS activity could predict 98.58% of the spiking activity at their targets. Only in 1.42% of the cases ΔTE took on negative values (Figure 9A, dark blue pixels), which shows that PS activity at individual sites could be predicted reversely from the spiking activity of their postsynaptic partners.

Further on we asked if the local effect is preserved at network level. In this case, the PS activity at the eight selected channels represented now the senders and the NSTs were considered the receivers. As Figure 9B shows, the trend that was observed at local sites was also found at network level. Again, until 17 DIV, spiking activity of NSTs could be poorly predicted by PS activity only (mean $\Delta TE = 0.4 \cdot 10^{-2}$ bits/s with $SD = 0.3 \cdot 10^{-2}$ bits/s). Only two channels (i.e. 31 and 42) could be identified in causing spiking activity in the NSTs. From 17 DIV to 29 DIV, the mean ΔTE significantly increased to $1.5 \cdot 10^{-2}$ bits/s with a SD = $0.2 \cdot 10^{-2}$ bits/s suggesting that PS activity at local sites increasingly drove network activity. However, from 30 DIV onward, PS activity increased only slightly (mean ΔTE = $1.7 \cdot 10^{-2}$ bits/s with a SD = $0.3 \cdot 10^{-2}$ bits/s), mostly due to the same trend as observed at local sites. That is, the same four channels (50% out of the eight selected channels) presented larger ΔTE values and thus strengthened their influence on the spiking activity of each NST. Moreover, at network level, ΔTE never took on negative values, suggesting that PS activity at selected local sites always predicted NST activity and not vice versa.

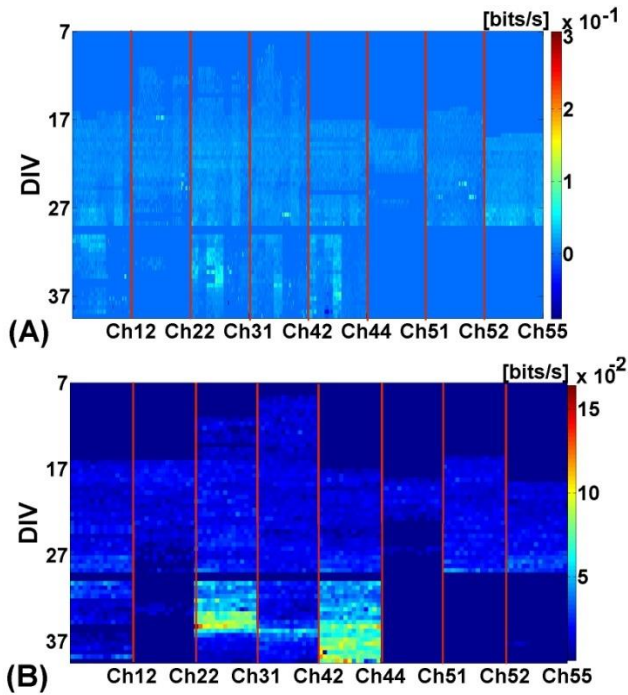

Figure 9 (**A**) Matrix of PS-related ΔTE (Methods 2.7) for eight recording sites. Vertical red bars delimit the sender channel that connects with the remaining seven receiver channels. Individual pixels represent the PS-related ΔTE averaged over a 30 minutes period and over the seven receiver channels. Each pixel row covers one DIV. (**B**) Matrix of PS-related ΔTE for eight senders (delimited by red bars) and the entire network as the receiver. Each pixel represents the ΔTE averaged over a 30 minutes period, each row covers one DIV.

## 4. Discussion

Neurons from dissociated brain tissue are capable of self-organizing their interconnectivity in cell culture. They become active even in the absence of any external sensory stimuli (Feller *et al.,* 1999). Besides random spiking and concerted bursting, neural networks use various modalities and activity patterns to both transfer information and form, as well as maintain, functional connections (Sun *et al.,* 2010). Such spontaneous, often synchronized neural activity increases in firing rate and in the number of active sites from one developmental stage to the next, as observed *in vivo* (Nadasdy 2000; Chiu *et al.,* 2001; Weliky *et al.,* 1999), as well as *in vitro* (van Pelt *et al.,* 2004, Wagenaar *et al.,* 2006; Rolston *et al.,* 2007; Pasquale *et al.,* 2010). Our un-interrupted long-term recording study, spanning over several weeks, confirmed this trend in two different neural *in vitro* networks, in a young hippocampal culture and for the first two periods in a more mature cortical culture. At later developmental stages, the activity and number of active sites slightly decreased. In addition, bursting frequency at network level did not increase anymore. Instead, it distributed spatially by involving more active sites. In contrast to previously reported snapshot activity recordings, our almost continuous recordings revealed large activity variations at particular DIVs. We therefore asked what network-inherent coding mechanisms shape and drive network activity.

Interestingly, we found that predominantly robust PS activity rather than bursts drove neural activity in the investigated cultures. PS not only developed stable spatiotemporal patterns, but also participated in shaping the interconnectivity map. Previous reports on its role *in vivo* and *in vitro* suggest that PS activity may act as a temporal filter and be part of a mechanism involved in information processing at different hierarchical information processing stages (Krahe 2004; Akerberg 2011; Rathbun *et al.,* 2010; Sincich *et al.,* 2009; Uglesich *et al.,* 2009).

In this study, the following main findings in support of these statements emerged from the analysis of the continuously recorded datasets:

1. The network firing rate, network burst rate and number of active sites increased as a hippocampal culture grew toward maturity and slightly decreased when the respective cultures started to decay.

2. At later developmental stages, spontaneous neural activity in a cortical culture oscillated periodically over several days. While activity could be very high at some DIVs, on average it evolved rather constantly.

3. PS activity became robust after three WIVs when IPSIs settled down to 2-3 s and the number of PS occurrences significantly increased at both individual sites and at network level. This may signal the passing of a critical maturation stage in spontaneously active *in vitro* networks.

4. In both cultures, highest PS activity could change its spatial location from one DIV to the next, which in most cases did not coincide with the location of the highest bursting activity. Furthermore, highest PS activity involved a larger number of neurons than the highest bursting activity, which stayed spatially confined to a few dominant electrodes throughout the entire recording period.

5. Although PS activity was found both outside and inside of bursting activity, it is an independent type of neural response, typically spatially separated from bursting activity. For the majority of the trials, the percentage of bursts which contained PS activity remained below 50%.

6. In most trials, PS activity preceded bursting activity at the same recording site with a lead of several tens of milliseconds. This suggests that PS may act as an internal surrogate stimulus, which triggers synchronized neural activity and avalanches (i.e. exemplified at 18 DIV, Figure 1C).

7. Its relation to other types of activity furthermore suggests that PSs act as network-intrinsic stimulus sources also at network level. From 11 DIV onward, network spiking activity was strongly correlated to the second spike of a PS with decreasing temporal delay as cultures matured.

8. Besides temporal patterns consisting of short IPSIs and frequent occurrence, PS activity formed increasingly complex spatial patterns during development by recruiting neighboring sites. Some of them fluctuated temporally; others lasted until the end of the recording session. A similar, yet activity type-unspecific increase in pattern complexity has been reported before (Rolston *et al.,* 2007; Sun *et al.,* 2010).

These findings combined with the above mentioned results suggest that PS activity evolves in distinct spatiotemporal patterns within a non-stimulated, spontaneously active network. PS presumably initiates synchronized bursting activity that may be responsible for forming particular functional connections both *in vivo* and *in vitro* (Rolston *et al.,* 2007; Sun *et al.,* 2010; Chiappalone *et al.,* 2012; Blankenship *et al.,* 2010; Mazzoni *et al.,* 2007; Pimashkin *et al.,* 2011).

9. CFP and information content per spike have been proven to reveal the statistical dependency between coupled neurons (Maccione *et al.*, 2012; le Feber *et al.*, 2007). Our CFP analysis revealed that PS activity is likely involved in the establishment and shaping of functional connections between different individual sites within the network. While we found few or no PS-related correlations between different sites at early stages when PS activity was not yet robust, the number of connections increased as the culture matured. They were characterized by rather constant average strengths and decreasing time delays for subsequent developmental stages. Interestingly, the PS-related information content per spike at individual sites was highest for the same period that the number of connections was found to be highest. These findings suggest a consolidation of PS-correlated activity at individual sites over time.

10. As the elevated PS-related information content per spike indicated, PS activity was involved in information transmission at individual sites and at network level. This finding is in concordance with previous studies on the role of different spiking patterns in spontaneous neural activity (Wagenaar *et al.,* 2006; Pasquale *et al.,* 2010; Sun *et al.,* 2010; Rolston *et al.,* 2007; Nadasdy 2000). It suggests that the network presumably uses only a fraction of the total number of spikes to transmit most of the information. As mentioned earlier, this concept of sparse coding was also found in the early visual system. It improves the overall coding efficiency by a mechanism that deletes the less informative spikes from one stage to the next (*i.e.* from the retina to the lateral geniculate nucleus (LGN)) while preserving relevant information with a lower number of spikes (Sincich *et al.,* 2009; Uglesich *et al.,* 2009).

11. The transfer entropy (Schreiber, 2000), an asymmetric information theoretic measure recently introduced to neuroscience, allows to estimate the direction of information flow within a network of locally coupled neurons (Ito *et al.*, 2011 – for spiking cortical network; Gourevitch and Eggermont, 2006 – for auditory cortical neurons; Garofalo *et al.*, 2009) or even between different brain areas (Battaglia *et al.*, 2012 – inter-areal brain circuits; Buehlmann and Deco, 2010; Lindner *et al.*, 2011 – directed interactions from the retina to the tectum; Lungarella and Sporns, 2006 – sensorimotor networks). Here we used TE to independently confirm that PS activity, besides forming spatio-temporal patterns and being involved in the formation of functional connectivity within the cultured neurons, also predicts the directionality of a connection at both local sites and network level (Figure 9A). TE analysis also strongly supported the trends revealed by information content per spike and CFP analysis. Presumably, PS activity was expansively involved in driving neuronal communication between 17 DIV and 29 DIV to arrive at a relative stability after 29 DIV where the PS neurons played a key role in driving neural activity in the already mature network.

The construction of NSTs by simply collecting the entire spiking activity and arranging the time stamps in an ascending temporal order may seem problematic at first glance. This way of constructing NSTs does not seem to have any biological relevance. However, using artificial spike trains and shuffling methods, we could show that PS in NSTs does not occur by chance. While artificial spike trains have similar firing rates as the recorded trains, no correlated activity between neurons (other than by pure random) is expected due to the lack of connectivity between individual neurons. The absence of PS-related connectivity for artificial spike trains suggests that PS activity is not a random and strictly firing rate-dependent neural phenomenon. Instead, it seems to be an intrinsic mechanism of cultured neurons in support of shaping neural interconnectivity. The local effect of PS activity seems to be preserved in NSTs as well. Figure 9B revealed a similar trend in the prediction of spiking activity at network level by PS. CFP analysis revealed a strong correlation between spiking activity at network level and each second spike in a PS. A similarity between CFP and information content per spike shapes strengthened the hypothesis that PS is involved in carrying information at network level as well. In summary, PS seems to play a key role in shaping the local and network-wide input-output relationship in cultured neural networks.

The two main hypotheses stated in this study have to be tested further in future experimental work. Firstly, in lack of any external stimulus, does PS indeed act as an internal surrogate stimulus that is capable of shaping neural activity by driving the input – output relationship of the spiking activity at network level? If this assumption turned out to be true, controlled PS-like electrical stimulation (similar to Zullo *et al.*, 2012) instead of single pulses or tetanic stimuli could more reliably drive a predictable neural output, *i.e.* in a closed-loop stimulation paradigm (Rolston *et al.,* 2010; Ruaro *et al.,* 2005; Novellino *et al.,* 2007). PS-like stimulation may find possible application in neurally-controlled artefacts (robotics, neuroprosthetics).

Secondly, because PS activity cannot only be found downstream of a stimulus both *in vitro* and in intact brain architectures (Usrey *et al.*, 1998; Sincich *et al.*, 2009), but also in spontaneously firing neural *in vitro* networks, it may represent an ubiquitous information processing response property of neurons of different species and organization. However, little is known on how PS depends on the neural cell type. A prominent example is the early visual system, where the PS contribution to information processing varies for different retinal ganglion cell types and the number of synaptic connections they are engaged in (Martiniuc *et al.*, in preparation). A combined electrophysiology and imaging *in vitro* study on how PS depends on cell type and cell morphology will shed more light on this question.

## Acknowledgements

We thank Francesca Succol and Marina Nanni for their expert advice and assistance in the cell culture preparation. IIT intramural funds in support of this research are highly appreciated.

# Supplementary Information

# Paired spiking robustly shapes spontaneous activity in neural networks *in vitro*

Aurel Vasile Martiniuc[1]*, Victor Bocoş-Binţinţan[2], Rouhollah Habibey[3], Asiyeh Golabchi[3], Alois Knoll[1], Axel Blau[3]

[1] Computer Science Department VI, Technical University Munich, Boltzmannstraße 3, 85748 Garching, Germany, http://www.in.tum.de/
[2] Dept. of Environmental Science & Engineering, Babeş-Bolyai University, 30 Fantanele Street, 400294 Cluj-Napoca, Romania, http://enviro.ubbcluj.ro/
[3] Dept. of Neuroscience and Brain Technologies, Italian Institute of Technology, via Morego 30, 16163 Genoa, Italy, http://www.iit.it

*Correspondence: martiniv@in.tum.de

## 1. Pictorial definition of spike train and burst parameters

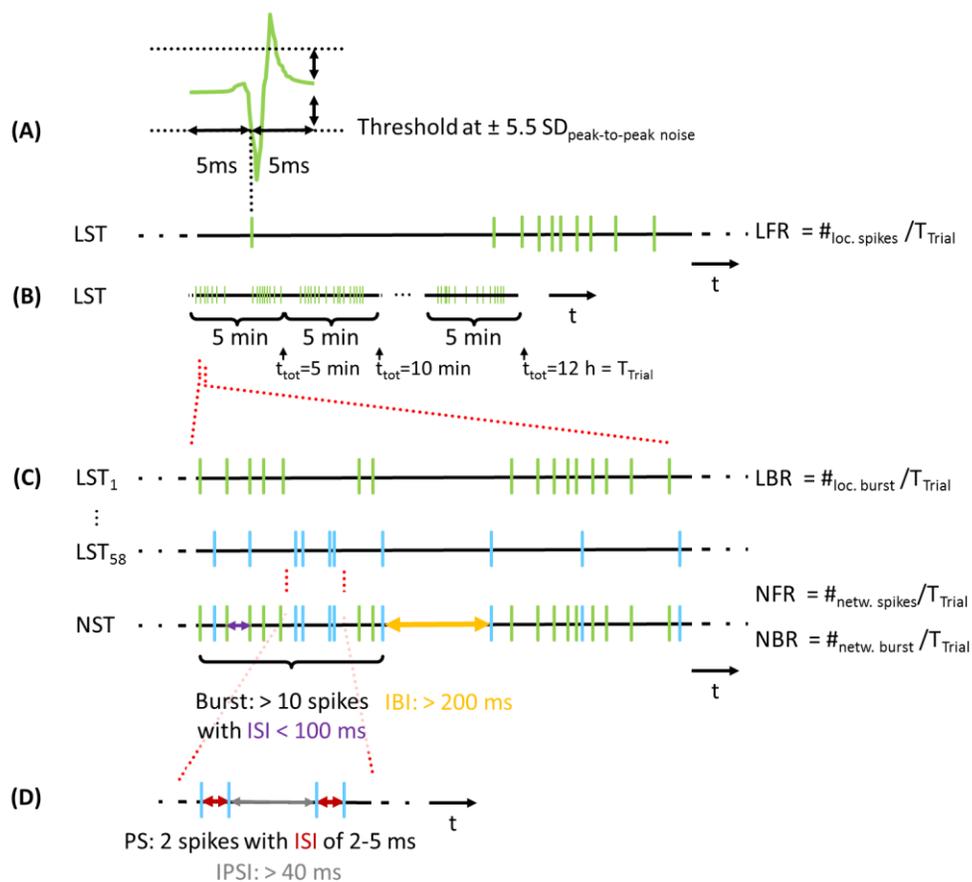

Suppl. Fig. 1    Graphical depiction of spike train and burst parameters. **(A)** Spike cutout: only upward (positive) and downward (negative) spike cutouts from 57 (cortical) and 58 (hippocampal) out of 59 recording electrodes were stored in 5 min packets. They consisted of 5 ms pre-spike and 5 ms post-spike fragments after first threshold crossing at ± 5.5 SD with respect to peak-to-peak noise. Only timestamps from downward threshold-crossings were extracted using

NeuroExplorer (Nex Technologies). After removing simultaneous timestamps that occurred on all channels due to electrical or handling artefacts, subsequent 5 min datasets comprised of ≤ 58 timestamp streams were bundled in 12 hour timestamp packets for further analysis in Matlab (MathWorks). These half-day packets were called trials of duration $T_{Trial} \leq 12$ h. **(B)** We quantified the local firing rates (LFRs) at individual sites as the number of recorded spikes divided by $T_{Trial}$ for each local spike train (LST). **(C)** At network level, we pooled all spikes from all 57 and 58 sites, respectively, for each trial into a single network spike train (NST) by sorting them in a time-ascending order. The NST represented the MEA-wide activity for each trial. The network firing rate (NFR) was then quantified as the total number of spikes in an NST divided by $T_{Trial}$. To further investigate activity dynamics, we used the burst rate (BR) for characterizing synchronous network activity. We scanned all LSTs at the 57 (cortical) and 58 (hippocampal) individual sites for each trial and defined bursting activity as events with more than 10 subsequent spikes individually separated by an ISI of less than 100 ms, followed by an interburst interval (IBI) larger than 200 ms (Wagenaar *et al.,* 2005). The local burst rate (LBR) at individual sites was calculated by dividing the number of bursts by $T_{Trial}$. Equally, the network burst rate (NBR) was obtained by scanning the NST for bursts using above mentioned criterion and dividing the number of bursts by $T_{Trial}$. **(D)** Definition of a paired spike. Two subsequent spikes with inter-spike intervals (ISIs) between 2 and 5 ms were considered paired spikes (PS). Only PS with inter-paired-spike intervals (IPSIs) over 40 ms were counted.

## 2. MEA layout

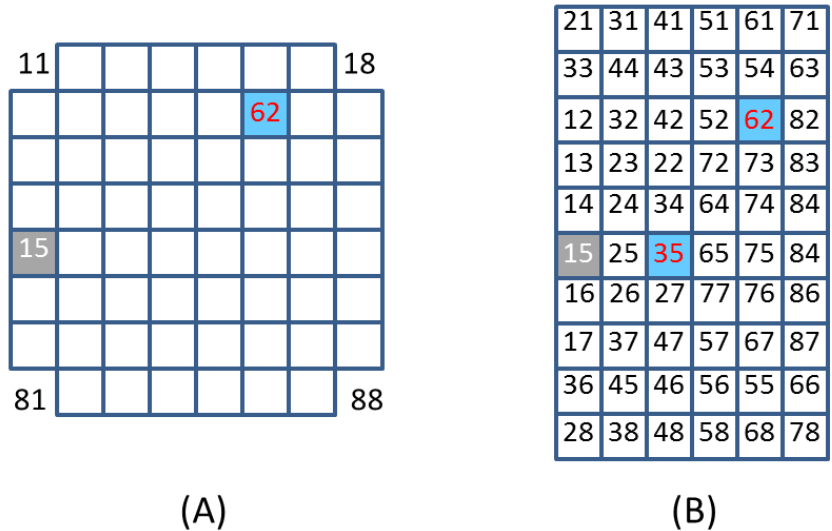

Suppl. Fig. 2   MEA electrode layout and channel association for the 8 x 8 MEA with the hippocampal **(A)** culture and for the 6 x 10 MEA with the cortical **(B)** culture. Grey squares indicate the relative position of the grounded counter electrode 15 (column, row). Blue squares with red numbers indicate switched-off channels.